\documentstyle[a4,11pt]{article}
\begin{document}

\begin{titlepage}
\vskip 2cm
\begin{flushright}
Preprint CNLP-1994-09
\end{flushright}
\vskip 2cm
\begin{center}
{\large {\bf On the Lakshmanan and gauge equivalent counterpart of the
Myrzakulov-VIII  equation}}\footnote{Preprint CNLP-1994-09.Alma-Ata.1994 \\
cnlpmyra@satsun.sci.kz}
\vskip 2cm

{\bf G.N.Nugmanova  }

\end{center}
\vskip 1cm

$^{a}$ Centre for Nonlinear Problems, PO Box 30, 480035, Almaty-35, Kazakstan

\begin{abstract}
The Lakshmanan equivalent counterparts of the some Myrzakulov equations
are found.
\end{abstract}


\end{titlepage}

\setcounter{page}{2}
\newpage

The structure of (1+1)-dimensional soliton equations is now very
well understood, and the question arises whether integrable
equations in (2+1)-dimension exist with such a structure. In the study of
nonlinear equations, a well know Lakshmanan and gauge equivalence (or
briefly, L-equivalence and g-equivalence respectively) take place between
the spin equations and the nonlinear Schrodinger-type equations [2-3].
In a series of papers [1] have studied families of new soliton equations
in (2+1)-dimensions. In this paper we find the Lakshmanan equivalent and
gauge equivalent counterpart of the M-VIII equation.

The so-called Myrzakulov-VIII equation (according to the notations of ref [1]) is
given by
$$ iS_t=\frac{1}{2}[S_{xx},S]+iu_xS_x   \eqno (1a) $$
$$ u_{xy}=\frac{1}{4i}tr(S[S_y,S_x])  \eqno (1b) $$
where the subscripts denote partial derivatives and S denotes the spin
matrix $ (r^2=\pm1)$

$$S= \pmatrix{
S_3 & rS^- \cr
rS^+ & -S_3
}, \eqno (2)$$
$$ S^2=I    $$

Eq. (1) is integrable,i.e. admits Lax representation [7] and different
type soliton solutions [8]. Now we wist find the Lakshmanan equivalent
counterpart of the M-VIII equation (1). To this end, starting from the results
of the ref.[8] we consider the
molion of sutface moving in
$ R^3 $ which generated by a position vector $\vec r(x,y,t)$. Here x and y
are local coordinates on the surface. The first and second fundamental
forms in the usual notation are given by
$$ I=d\vec r d\vec r=Edx^2+2Fdxdy+Gdy^2 \eqno (4) $$
$$ II=-d\vec r d\vec n=Ldx^2+2Mdxdy+Ndy^2 \eqno (5) $$
where $$ E=\vec r_x\vec r_x=g_{11}, F=\vec r_x\vec r_y=g_{12},
G=\vec r_{y^2}=g_{22},  \eqno (6) $$
$$ L=\vec n\vec r_{xx}=b_{11}, M=\vec n\vec r_{xy}=b_{12},
N=\vec n\vec r_{yy}=b_{22}, \vec n=\frac{(\vec r_x \wedge \vec r_y)}
{|\vec r_x \wedge \vec r_y|} \eqno (7) $$

At a fixed time, the tangent  $ (\vec r_x, \vec r_y) $ and normal
$ (\vec n) $ vectors satisfy the following Gauss-Weingarten equations (GUE)
$$ \vec r_{xx}=\Gamma^1_{11} \vec r_x + \Gamma^2_{11} \vec r_y +L \vec n
\eqno (8a) $$
$$ \vec r_{xy}=\Gamma^1_{12} \vec r_x + \Gamma^2_{12} \vec r_y +M \vec n
\eqno (8b)$$
$$ \vec r_{yy}=\Gamma^1_{22} \vec r_x + \Gamma^2_{22} \vec r_y + N \vec n
\eqno (8c)$$
$$ \vec n_x=p_1 \vec r_x+p_2 \vec r_y \eqno (8d) $$
$$ \vec n_y=q_1 \vec r_x+q_2 \vec r_y \eqno (8e) $$
where $ \Gamma^k_{ij} $ are the Christoffel symbols of the second kind defined by
the metric $ g_{ij} $ and $ g^{ij}=(g_{ij})^1 $ as
$$ \Gamma^k_{ij}=\frac{1}{2} g^{kl}(\frac{\partial g_{lg}} {\partial x^i}+
   \frac {\partial g_{il}}{ \partial x^j}-\frac{\partial g_{ij}}
   {\partial x^l}) \eqno (9) $$
The coefficients $ p_i, q_i $ are given by
$$ p_i=-b_{1j}g^{ji} \eqno (10a) $$
$$ q_i=-b_{2j}g^{ji} \eqno (10b) $$
The compatibility conditions $ \vec r_{xxy}=\vec r_{xyx} $ and
$ \vec r_{yyx}=\vec r_{xyy} $ yield the following Mainardi-Peterson-Codazzi
equations (MPCE)
$$ \frac{\partial \Gamma^l_{ij}}{\partial x^k}-\frac{\partial \Gamma^l_{ik}}
{\partial x^j}+\Gamma^s_{ij} \Gamma^l_{ks}-\Gamma^s_{ik} \Gamma^l_{js}=
b_{ij}b^l_{k}-b_{ik}b^l_{j}.
\eqno (11a) $$
$$ \frac{\partial b_{ij}}{\partial x^k}-\frac{\partial b_{ik}}{\partial
    x^j}=\Gamma^s_{ik}b_{is}-\Gamma^s_{ij}b_{ks}
\eqno (11b) $$
where $ b^j_i=g^{jl}b_{il}, x^1=x,x^2=y. $

Let $ E=\vec r_x^2=1$. Then we can introduce the orthogonal trihedral - the
new basic unit vectors [1]
$$ \vec e_1=\vec r_x, \vec e_2= \vec n, \,\,\
\vec e_3=\vec e_1\wedge \vec e_2.
 \eqno (12) $$

Now from () and () after some algebra follows that the vectors $ \vec e_i $
satisfy the following set of equations
$$ \vec e_{ix}=\vec D \wedge \vec e_i \eqno (13a) $$
$$ \vec e_{iy}=\vec M \wedge \vec e_i \eqno (13b) $$
where $$ \vec D= \tau \vec e_1+k \vec e_3,  \vec M=m_1 \vec e_1+m_2 \vec e_2
+ m_3 \vec e_3,   \,\,\, k = L, \,\,\,
\tau=\frac{L^2N}{\sqrt G} $$
$$(m_1,m_2,m_3)=     $$
Motion of a surface in the basic $ \vec e_i $ can be prescribed in the form
$$ \vec e_{it} =\vec\Omega\wedge \vec e_i,
\,\,\,\, \vec\Omega=\sum_{j=1}^{3}w_j \vec e_j \eqno(13c)   $$
where $ w_j(x,y,t) $ are temporally unknown functions. Now let $\vec e_1=
\vec S, $ where $ \vec S $ is the spin vector satisfing the vector form
of equation (1)
$$ \vec S_t + \vec S\wedge \vec S_{xx} + u_x\vec S_x = 0 \eqno(14a)$$
$$ u_{xy} = -\vec S(\vec S_x \wedge\vec S_y) \eqno(14b) $$
With these choices, the functions $ w_j $ take the following expressions
$$ w_3=k(\tau -u_x) $$
$$ w_2=-k_x $$
$$ w_3= \frac{k_{xx}}{k}+\tau^2-u_x\tau $$

Now we ready to find the L-equivalent counterpart of equation  (1)(or(14)).
Let us introduce the new complex function
$$ \psi (x,y,t)=ae^{ib} $$
where $ a(x,y,t)$ and $b(x,y,t)$ are the some function of $\vec S $ [1],
for example,
$$
4a^{2}(x,y,t) = \alpha (2a + 1)\vec S_{x}\vec S_{y} +
i\alpha \vec S(\vec S_{x}\wedge \vec S_{y})
- a(a+1) \vec S^{2}_{x} -\alpha^{2}\vec S_{y}^{2}
.\eqno(17) $$
Then $ \psi(x,y,t) $ satisfies the following Zakharov equation [4]
$$ iq_t + q_{xx} + v q = 0,  \eqno(18a) $$
$$ ip_t - q_{xx} - v q = 0,\eqno(18b) $$
$$ v_y=2(pq)_x \eqno(18c) $$
where $p = r^{2}q$. This set of equationis the equivalent counterpart
of eqs.(1).
On the other
hand,in [5] was shown that eqs.(1) and (18) are gauge equivalense each other.

Finally, we note that eqs.(1) are the particular case of the Myrzakulov-IX equation
[1]
$$ iS_t+\frac{1}{2}[S,M_1S]+A_1S_x+A_2S_y \eqno(19a)$$
$$ M_2u=\frac{\alpha}{4i}tr(S[S_y,S_x]) \eqno(19b)$$
where $ \alpha,b,a  $ consts and
$$ M_1= \alpha ^2\frac{\partial ^2}{\partial y^2}-2\alpha (b-a)\frac{\partial^2}
   {\partial x \partial y}+(a^2-2ab-b)\frac{\partial^2}{\partial x^2}; $$
$$ M_2=\alpha^2\frac{\partial^2}{\partial y^2} -\alpha(2a+1)\frac{\partial^2}
   {\partial x \partial y}+a(a+1)\frac{\partial^2}{\partial x^2},$$
$$ A_1=i\alpha\{(2ab+a+b)u_x-(2b+1)\alpha u_y\} $$
$$ A_2=i\{\alpha(2ab+a+b)u_y-(2a^2b+a^2+2ab+b)u_x\}, $$
Eqs.(19) admit the two integrable reductions. As b=0, eqs. (19) after the
simply transformation reduces to the M-VIII equation (1) and as
 $ a=b=-\frac{1}{2} $
to the Ishimori equation. In general we have the two integrable cases:
 the M-IXA equation as $\alpha^{2} = 1,$ the M-IXB equation
as $\alpha^{2} = -1$. We note that the M-IX equation is integrable and
admits the Lax representation [1]
$$ \alpha \Phi_y =\frac{1}{2}[S+(2a+1)I]\Phi_x \eqno(20a) $$
$$ \Phi_t=\frac{i}{2}[S+(2b+1)I]"_{xx}+\frac{i}{2}W\Phi_x \eqno(20b) $$
where $$ W_1=W-W_2=(2b+1)E+(2b-a+\frac{1}{2})SS_x+(2b+1)FS $$
$$ W_2=W-W_1=FI+\frac{1}{2}S_x+ES+\alpha SS_y $$
$$ E = -\frac{i}{2\alpha} u_x,\,\,\,  F = \frac{i}{2}(\frac{(2l+1)u_{x}}{\alpha} -
2u_{y}) $$

Hence as $b = 0$ we get the Lax refresentation of the M-VIII(1). Both
the M-IX and M-VIII equations  admit the different type exact solutions
(solitons, lumps, vortex-like, dromion-like and so on)[7].
As shown in [6] eqs. (19) have the following L-equivalent
and g-equivalent[5]) counterpart
$$ iq_t+M_1q+vq=0 \eqno(21a)$$
$$ iq_t-M_1p-vp-0 \eqno(21b)$$
$$ M_2v=M_1(pq) \eqno(21c)$$
where $p=r^{2}\bar q$.
As well known these equations are too integrable [4] and as in the previous
case, eqs. (21) have the two integrable reductions:eqs(18)as b=0 and the
Davey-Stewartson equation as $ a=b=-\frac{1}{2}. $
\\
\\
{\bf APPENDIX}
\\
Here we present the L- and g- equivalent of the M-XXII equation
$$ -iS_t=\frac{1}{2}([S,S_y]+2iuS)_x+\frac{i}{2}V_1S_x-2ia^2 S_y $$
$$ u_x=-\vec S(\vec S_x\wedge \vec S_y) $$
$$ V_{1x}=\frac{1}{4a^2}(\vec S^2_x)_y $$
It has the form
$$ q_t=iq_{yx}-\frac{1}{2}[(V_1q)_x-qV_2-qrq_y] $$
$$ r_t=-ir_{yx}-\frac{1}{2}[(V_1r)_x-qrr_y+rV_2] $$
$$ V_{1x}=(qr)_y $$
$$ V_{2x}=r_{yx}q-rq_{yx} $$
Both these set of equations are integrable and the corressponding
Lax representations were presented in[1].

\end{document}